\documentstyle[12pt]{article}
\newcommand{\be}{\begin{equation}}
\newcommand{\ee}{\end{equation}}
\newcommand{\ba}{\begin{eqnarray}}
\newcommand{\ea}{\end{eqnarray}}
\newcommand{\baa}{\begin{eqnarray*}}
\newcommand{\eaa}{\end{eqnarray*}}
\newcommand{\bb}{\begin{thebibliography}}
\newcommand{\eb}{\end{thebibliography}}
\newcommand{\ci}[1]{\cite{#1}}
\newcommand{\bi}[1]{\bibitem{#1}}
\newcommand{\lab}[1]{\label{#1}}

\begin{document}
\begin{flushright}
JINR preprint E2-94-262, Dubna, 1994
\end{flushright}
\vspace{15mm}

\begin{center}
{\large Spin Effects in pp-scattering
         at difraction range and  RHIC energies} \\ [10mm]
 S.V.Goloskokov$^1$, O.V.Selyugin$^2$ \\
\vspace{.5cm}
Bogolubov Laboratory of Theoretical Physics,\\
Joint Institute for Nuclear Research, Dubna\\
Head Post Office P.O.Box 79, 101000 Moscow, Russia
\end{center}

\begin{abstract}
   The spin effects in nucleon-nucleon scattering are calculated in
the framework of  the
dynamic model  taking  into  account  interactions  at  large
distances and the strong form  factors; the model factorization
of the $NN$ amplitude into the spin-dependent hadron-pomeron vertex
and high-energy spinless pomeron is obtained.
Theoretical predictions for polarization and $A_{nn}$  are made
for the diffraction range at the RHIC energies.
\end{abstract}

\vspace{2cm}
\begin{flushleft}
------------\\
$^{1} $ E-mail: goloskkv@thsun1.jinr.dubna.su;
\end{flushleft}
\begin{flushleft}
$^{2} $ E-mail: selugin@thsun1.jinr.dubna.su
\end{flushleft}

\newpage
\phantom{.}
\vspace {2cm}

       Some important results are obtained at the CERN collider $S\bar p pS$
and FNAL in investigating elastic scattering.
 It is just this process that allows the verification of the
results obtained from the main principles of quantum field theory:
the concept of the scattering amplitude as a unified analytic function
of its kinematic variables connecting different reaction channels
were introduced in the dispersion theory by N.N. Bogolubov\ci{bog}.
The results of $\bar{p} p$- collider show a still continuing growth of the
total cross section, the diffraction peak shrinkage and the slow growth
of the relation of $\sigma_{elastic}/\sigma_{tot}$. Especial question
is about the behavior of $\rho = Re T(s,0)/Im T(s,0)$, which
is tightly connected
with the dispersion relation. The large magnitude of $\rho$
measured by the UA4 Collaboration \ci{ua4} is in contradiction
with the first analysis of the UA4/2
Collaboration \ci{UA4/2}, but was confirmed by the next one \ci{sel}.
     In most of the early models, as in the ordinary picture
of PQCD, the spin effects were suppressed at these energies.
However, in some models \ci{Kane}, the spin-flip amplitudes which don't
decrease with the growing energy, were predicted. For example, in the
model \ci{ETT} the absence of the second diffraction minimum was
explained by the spin-flip contributions. In framework of the QCD,
spin effects weakly dependent on energy were
shown to exist by analysing the spin structure of a quark-pomeron vertex
\ci{gol1}. Now, some different
models examining the nonperturbative instanton contribution lead
to sufficiently large spin effects at superhigh energies \ci{Forte},
\ci{Dor}.
Careful analyses of experimental data also
 show a possible of manifestation spin-flip amplitudes
at high energies \ci{sel,akch}. The research of such spin effects
will be a crucial stone for different models and will help us
to understand the interaction and structure of particles, especially at large
distances.
All this raises the question about the measure of spin effects
in the elastic hadron scattering at small angles  at
future accelerator ( HERA, RHIC, LHC and UNK).
Now, there are large programs of reserching spin effect
at these accelerators. Especially, we should like to note
the programs at RHIC \ci{rhic},
where the polarization of both the collider beams will be constructed.
So, it is very important to obtain reliable predictions for the spin
assymmetries at large energies. In this paper, we extend the model
predictions for the spin asymmetries in the RHIC energy domain.
The factorization of the scattering amplitude into the spin-dependent
hadron-pomeron vertex function and high energy pomeron is shown
in the model too.

     In papers \ci{gks1,gks}, the dynamical model for a particle interaction
which takes into account the hadron structure at large distances
was developed.
The model is based on the general quantum field theory principles
(analyticity, unitarity and so on) and takes into account basic information
on the structure of a hadron as a compound system with the central part region
where the valence quarks are concentrated and the long-distance region
where the color-singlet quark-gluon field occurs.
 As a result, the
hadron amplitude can be  represented  as  a  sum  of the central  and
peripheral parts of the interaction \ci{g2}:
\ba
T(s,t) \propto T_{c}(s,t) + T_{p}(s,t).  \lab{tt}
\ea
 where $T_{c}(s,t)$   describes   the
interaction between the central parts of hadrons.
At high energies it is determined by the spinless pomeron exchange.
The   $T_{p}(s,t)$  is
the sum of triangle diagrams corresponding to
the interactions of the central part of one hadron  on  the  meson
cloud of the other. The meson - nucleon interaction leads to the spin flip
effects in the pomeron-hadron vertex.

          The contribution of these triangle diagrams
to the scattering amplitude  with  $N(\Delta $-isobar)  in  the
intermediate state looks like as follows \ci{zpc}:

\ba
T ^{\lambda _{1}\lambda _{2}}_{N(\Delta )}(s,t) =
{g^{2}_{\pi NN(\Delta )} \over i(2\pi )^{4}}
  \int  d^{4} q T_{\pi N} (s^{\prime}, t) \varphi_{N (\Delta )}
[(k-q),q^{2}] \varphi_{N(\Delta )} [(p-q),q^{2}]    \nonumber \\
\times \frac{\Gamma^{\lambda_{1}\lambda_{2}}(q,p,k,)}
{
[q^{2} - M^{2}_{N (\Delta) } + i \epsilon ] [(k-q)^{2} - \mu^{2} +
 i \epsilon ]  [(p-q)^{2} - \mu^{2} + i \epsilon ]
}.   \lab{tint}
\ea
Here $\lambda_{1},\lambda_{2}$  are  helicities  of  nucleons;
$T_{\pi N}$  is  the
$\pi N$-scattering amplitude; $\Gamma$  is a matrix element of the numerator
of the diagram ; $\varphi $ are vertex functions chosen
 in the dipole form  with
the parameters $\beta _{N(\Delta )}$:
\ba
\varphi _{N(\Delta )}(l^{2},q^{2}\propto  M^{2}_{N(\Delta )})
  = {b^{4}_{N(\Delta )}\over
      (b^{2}_{N(\Delta )}- l^{2})^{2}}.
   \lab{fi}
\ea

       For a standard form of the pomeron contribution to the
meson-nucleon scattering amplitude
$$
    T_{\pi N} (s,t) = i \beta^{\pi}(t) \cdot \beta^{N}(t) s^{\alpha(t)}
$$
we can write the integral (\ref{tint}) in the form:
$$
    T^{\lambda_1 \lambda_2}_{N (\Delta)}(s,t) =
  i \beta^{N(\lambda_1 \lambda_2)}(t) \cdot \beta^{N}(t) s^{\alpha(t)},
$$
where  $\beta^{N(\lambda_1 \lambda_2)}(t)$ is a $N$-nucleon or
$\Delta_{33}$-isobar contribution to a spin-dependent nucleon-pomeron
vertex function. In the light-cone variables
$q=(x p_{+}, q_{-}, q_{\perp})$,$ q_{\pm}=q_{0} \pm q_{z}$;
it has the form:
\ba
 \beta^{N(\lambda_1 \lambda_2)}=
{\beta^{\pi}(t) g^{2}_{\pi NN(\Delta )} b^{8}_{N(\Delta)}\over 2(2\pi )^{3}}
  \int_{0}^{1}
 \!\! dx x^5 (1-x)^{\alpha(t)}
\!\!\!\int \!\!\frac
{    d^2 \vec{q}_{\perp} \Gamma^{\lambda_{1}\lambda_{2}}_{N(\Delta)}
(\vec{q_{\perp}},\vec{p},\vec{k},)}
{[q^{2}_{\perp}+d][q^{\prime 2}_{\perp}+d]
[q^{2}_{\perp}+a]^2 [q^{\prime 2}_{\perp}+a]^2}.   \lab{int2}
\ea
Here
\ba
\vec{q}\,^{\prime}=\vec{q}_{\perp}+x(\vec{p}-\vec{k});
  d=(M^{2}_{N(\Delta)}-xM^{2}_{N})(1-x)+\mu^2 x;  \nonumber \\
  a=(M^{2}_{N(\Delta)}-xM^{2}_{N})(1-x)+b^{2}_{N(\Delta)} x. \nonumber
\ea
As a consequence we obtain the factorization of the scattering
amplitude into the spin-dependent hadron-pomeron vertex function and high
energy pomeron, as it has been obtained early in the frame work of QCD
\ci{gol1}.

The  matrix  element  of  the  nucleon-intermediate-state
contribution has the form:
\ba
\Gamma^{++}_{N}= [ M^{2}_{N} (1 - x)^{2}/x + q^{2}_{\perp}/x
- \vec{q}_{\perp}\vec{\Delta }_{\perp}];       \lab{gpp}
\Gamma^{+-}_{N}= \Delta M_{N} (x-1);
\ea
 where $\Delta $ is the transfer momentum. From (\ref{gpp})
it is  seen  that  the spin-flip and spin-non-flip amplitudes
have the  same  asymptotics in $s$.

     The matrix element of the $\Delta _{33}$-isobar contribution
in the intermediate state has the form:
\ba
\Gamma^{\lambda_{1}\lambda_{2}}_{\Delta }= \bar{u}(p)^{\lambda_1}
(\hat{q}+M_{\Delta})[(pk) - {1\over 3} \hat{p}\hat{k} -
{2(pq)(kq\over 3 M^{2}_{\Delta }}^{)}+ {(pq)\hat{k} - (kq)
\hat{p}\over 3 M_{\Delta }}] u(k)^{\lambda_2}.   \lab{gll}
\ea
 The spin-flip and no-flip matrix element can be found in \ci{zpc}.

      The consideration of isotopic factors in integrals (\ref{tint})  leads
to the following expressions for the amplitude of the $NN$ interaction:
\ba
T^{+-}=3T^{+-}_{N} + 2 T^{+-}_{\Delta } . \lab{tnd}
\ea
  Different signs of the amplitudes (\ref{tnd}) essentially  compensate  the
contributions of $N$  and $\Delta $  states  with  each  other  in  elastic
processes.
The dipole parametrization for meson-baryon vertex functions
is used with the following values of the parameters \ci{lom}:
$$
b^{2}_{N} =  3.4(GeV^{2}), b^{2}_{\Delta }= 1.5(GeV^{2})
$$
 and coupling constants \ci{mach}:
$$
{
g^{2}_{\pi N N} \over 4\pi
}
= 14.6;
{
g^{2}_{\pi N \Delta} \over 4\pi
}
=21 (GeV^{2}).
$$
    Note that in integrals (\ref{tint}) the $X \simeq 0.9$ range
is  essential,
which makes it necessary a correct consideration of the contribution of a
sufficiently-low-energy range $s^{\prime}\simeq 0.1s$. For this,
in  calculating
integrals (\ref{tint}) we use the spin-non-flip amplitude obtained in
\ci{g13} with the $1/\sqrt{s}$  terms which describes the  experimental  data
of
meson-nucleon scattering in a wide energy range.
     The  peripheral
contribution calculated in the model leads to the  spin  effects
in the Born term of the scattering amplitude which do not disappear
with growing energy.
 Summation of rescattering at s-channel has been performed with the help
of the quasipotential equation. The total amplitude has an eikonal form.
The explicit forms of the helicity amplitudes and the parameters
obtained can be found in \ci{zpc}.

        The model with  the $N $  and  $\Delta $
contribution  provides  a self-consistent  picture  of
the differential cross sections and spin phenomena
of different hadron processes  at  high  energies.
Really, the parameters in the  amplitude  determined from one
reaction, for example, elastic $pp$-scattering, allow one to  obtain
a wide range of results for elastic
meson-nucleon scattering and charge-exchange reaction
 $\pi^{-} p \rightarrow  \pi^{0} n$
 at high energies.

     The model predicts that at superhigh energies  the  polarization
effects of particles and antiparticles are the same.
Though the polarization of the proton-antiproton scattering at
$ \sqrt{s}=16.8 GeV$ is larger than the proton-proton polarization ,
they become practically equal at energy
$\sqrt{s} \geq 30 GeV$. After $\sqrt{s}=50 GeV$, the polarization
decreases very slowly and has the very determining form.
It is small at small transfer momenta, before the diffraction peak,
and has the narrow, sufficiently large negative peak in the range
of the diffraction minimum, see Fig.1.

          The magnitude of the negative peak slowly falls down
>from 0.52 at $\sqrt{s}=50 GeV$ up to 0.25  at $\sqrt{s}=500 GeV$.
It has the width at half the high   $\Delta(t)= 0.17 GeV^2$
at $\sqrt{s}=50 GeV$  and nearly
constant $\Delta(t)= 0.3 GeV^2$ for  $\sqrt{s} > 150 GeV$.
The position of the negative maximum slowly shift to small transfer momenta,
see Table 1.
Behind the diffraction minimum the polarization
changes its sign and has the bump up to $|t|=6 GeV^2$.
The position of maximum of this bump slowly changes towards larger $|t|$
and its magnitude some what change around $10\%$;
near $|t|=3 GeV^2$ the magnitude is practically
the constant $=8\%$ (see Table 1).

    Our predictions coincide very closely with the results
of the model \ci{BS}
at sufficiently low energies ($\sqrt{s} \simeq 23.4 GeV$).
But these models lead to very different predictions at superhigh (RHIC)
energies. Especially, it concerns the range of the diffraction
minimum. So, if our model leads to the negative polarization,
the model \ci{BS} gives the positive polarization.

      The behavior of the spin correlation parameter $A_{NN}$ is  shown
in fig.2.
The magnitude of $A_{nn}$, its width at half the height - $\Delta(t)$,
the magnitude $A_{nn}$ at $|t|=3 GeV$ and in the second maximum
are shown in Table 2.
As is seen, the value of $A_{NN}$  becomes  maximum  in  the
range of the  diffraction  minimum,  as  in  the  case  of
the polarization.
The magnitude of $A_{nn}$ become sufficiently large with the growth of $|t|$.
The reason is that in this work we have used
the strong form factors for the vertices $\pi N N $  and $\pi N\Delta $.
The form of the spin-flip amplitude is determined in
the model up to $\mid t\mid  \simeq  2$ GeV$^{2}$, hence,  we  can
expect  an  adequate
description of the experimental data up to
$\mid t\mid  \simeq  3 \div  5.0$  GeV$^{2}$.

       It should be emphasized that  the  model  gives  the  same  energy
dependence  of  the  spin-flip  and  spin-non-flip   amplitudes.
Consequently,  the obtained spin  effects  do not
disappear in the asymptotic energy range. The model predicts large
values of the polarization  and  spin  correlation  parameter $A_{NN}$
which decrease slowly from
$22\%$  to $6.4\%$  at the range of the diffraction minimum  with
increasing energy from $\sqrt{s}=50 GeV$ to
$\sqrt{s}=500 GeV$. Note that  the
polarizations of $pp$-  and $p\bar{p}$-  scattering  will  coincide  above
the energies $\sqrt{s}> 30 GeV$.
Note that in our model the standard helicity amplitude $\phi_{1}$
equals $\phi_{3}$
and $\phi_{2}$, $\phi_{4}$ are an order as small as
$\phi_{1}$ and have kinematical factor $f(t)$.
Hence $\delta\sigma_{L}$  equals zero and
$\delta\sigma_{T}$ is practically invisible in the domain before
diffraction minimum.

    Thus, the dynamical model considered, which takes into account
 the $N$ and $\Delta$ contribution, leads
to a lot of predictions concerning
the behavior of spin correlation parameters at high energies.
In that model, the effects of large  distances
determined  by  the  meson  cloud  of  hadrons  give  a   dominant
contribution to the spin-flip amplitudes  of  different  exclusive
processes at high energies and fixed transfer momenta.
Note that the results on the spin effects obtained here differ from
the predictions of other models \ci{BS} at an energy above
$\sqrt{s} \geq 30 GeV$.
And the examination of these results gives new information about the
hadron interaction at large distances.

The   authors   express   their   gratitude   to  S.P.Kuleshov,
 V.A.Meshcheryakov, A.N.Sis-sakian for the support
in this work and to N.Akchurin, A.V.Efremov, W.D.Nowak, S.B.Nurushev,
J.Soffer
for the discussion of the problems of the measure of the spin effects
at RHIC and HERA energies.

    This work was supported in part by the Russian Fond of
Fundamental Research, Grand $   94-02-04616$.

\newpage

\begin{table}
\begin{tabular}{|c|c|c|c|c|c|} \hline
	&               &            &             &               &  \\
$\sqrt{s}$& $|t|_{max}$ & $P_{1max}$ & $\Delta(t)$ & $P(|t|=3GeV^2)$&
$P_{2max}$ \\
  $GeV$ & $GeV^2$   & $\% $ & $GeV^2$     &     $\%$ & \%  \\ \hline
	&           &       &            &    & \\
 $50$   & 1.337     & -52  &  0.17    &   10.7 & 12.3 \\
	&           &       &           &          &         \\
 $100$   & 1.225     & -41  &  0.26    &    8.5 &  9.2 \\
	&           &       &          &&  \\
 $150$   & 1.15      & -37  &  0.28    &    7.9 &  8.6 \\
	&           &       &           && \\
 $200$   & 1.09      & -34  &  0.29    &    7.8 &  8.6 \\
	&           &       &           && \\
 $250$   & 1.05      & -32  &  0.30    &    7.7 &  9.1 \\
	&           &       &            && \\
 $300$   & 1.005     & -30  &  0.30    &    7.7 & 10.1 \\
	&           &       &           && \\
 $350$   & 0.98      & -28  &  0.30    &    7.7 & 11.2 \\
	&           &       &           && \\
 $400$   & 0.95      & -27  &  0.30    &    7.7 & 12.8 \\
	&           &       &           && \\
 $450$   & 0.925     & -26  &  0.29    &    7.7 & 14.3 \\
	&           &       &            && \\
 $500$   & 0.905     & -24  &  0.29    &    7.7 & 16.0 \\
	&           &       &            &  & \\  \hline
\end{tabular}
\caption{The theoretical predictions of the polarization at RHIC
energies }
\end{table}

\newpage

\begin{table}
\begin{tabular}{|c|c|c|c|c|c|} \hline
	&               &            &             &               &  \\
$\sqrt{s}$& $|t|_{max}$ & $A_{nn}^{1max}$ & $\Delta(t)$ & $A_{nn}(|t|=3GeV^2)$&
$A_{nn}^{2max}$ \\
  $GeV$ & $GeV^2$   & $\% $ & $GeV^2$    &     $\%$ & \% \\ \hline
	&           &       &            &    & \\
 $50$   & 1.30     &  22  &  0.27    &   10.7 & 11.7 \\
	&           &       &           &          &         \\
 $100$   & 1.15      & 16.4 &  0.33    &    7.6 &  8.8 \\
	&           &       &          &&  \\
 $150$   & 1.08      & 13.6 &  0.33    &    6.2 &  7.5 \\
	&           &       &           && \\
 $200$   & 1.0       & 11.7 &  0.32    &    5.3 &  6.6 \\
	&           &       &           && \\
 $250$   & 0.95      & 10.3 &  0.32    &    4.7 &  5.9 \\
	&           &       &            && \\
 $300$   & 0.93      &  9.3 &  0.32    &    4.2 &  5.4 \\
	&           &       &           && \\
 $350$   & 0.92      &  8.5 &  0.32    &    3.8 &  4.9 \\
	&           &       &           && \\
 $400$   & 0.90      &  7.6 &  0.32    &    3.4 &  4.5 \\
	&           &       &           && \\
 $450$   & 0.87      &  7.1 &  0.31    &    3.1 &  4.2 \\
	&           &       &            && \\
 $500$   & 0.85      &  6.4 &  0.28    &    2.9 &  3.9 \\
	&           &       &            &  & \\  \hline
\end{tabular}
\caption{The theoretical predictions of the $A_{nn}$ at RHIC
energies }
\end{table}

\newpage
\phantom{.}
\vspace{2cm}
\begin{center}
{\large                         Figure captions     } \\
\end{center}
 Fig. 1 The predictions for Polarization of the $ p p$ - scattering  \\
\phantom{.} \hspace{1.5cm} at $\sqrt{s}=23.4 GeV$
and the experimental points \ci{kl}.   \\

    Fig. 2 The predictions for Polarization of the $ p p$ - scattering
at RHIC energies ; \\
\phantom{.} \hspace{1.5cm}
solid curve - at $\sqrt{s}=50 GeV$ and
dashed curve - at $\sqrt{s}=300 GeV$.  \\

 Fig. 3 The predictions for the $A_{nn}$ at RHIC energies.  \\
\phantom{.} \hspace{1.5cm}
solid curve - at $\sqrt{s}=50 GeV$ and
dashed curve - at $\sqrt{s}=300 GeV$.
\end{document}